\documentclass[12pt]{article} 
\title{A Derivation of Vector and Momentum Matrices}  
\author{{\it Richard Shurtleff~}\thanks{affiliation and mailing 
address: Department of Applied Mathematics and Sciences, 
Wentworth Institute of Technology, 550 Huntington Avenue, 
Boston, MA, USA, ZIP 02115, telephone number: (617) 989-4338, e-mail address: shurtleffr@wit.edu}} 
\begin{document} 

\maketitle 
              
\begin{abstract} 

Given standard angular momentum and boost matrices, the commutation rules for vector and momentum matrices are solved. The resulting matrix components are displayed as detailed functions of spin with factors such as $ \sqrt{2A+1}.$ For comparison and as an alternative, Lyubarskii's formulas in terms of Clebsch-Gordan coefficients are recalled from the literature and displayed. A set of these momentum matrices combined with the corresponding set of six angular momentum and boost matrices form the generators of a nonunitary  finite dimensional representation of the Poincar\'{e} group of translations, rotations and boosts. A problem set is included.

\vspace{0.5cm}


\end{abstract}


\section{Introduction } \label{intro}  

A vector matrix is both a vector  and a tensor. Under a Lorentz transformation ${\Lambda_{\mu}}^{\nu}$ the vector matrix $V_{\mu ij}$ transforms like a vector: $V_{\mu ij} \rightarrow$  ${\Lambda_{\mu}}^{\nu}V_{\nu ij}.$ Also, given an $n$-dimensional representation $D_{ij}(\Lambda)$ of the proper homogeneous Lorentz group, the vector matrix transforms as a second rank tensor:  $V_{\mu ij} \rightarrow $ $D_{ij}(\Lambda) V_{\mu jk} D^{-1}_{kl}(\Lambda).$ 
Equating the results, we get
\begin{equation} \label{DvD} D_{ij}(\Lambda) V_{\mu jk} D^{-1}_{kl}(\Lambda) = {\Lambda_{\mu}}^{\nu}V_{\nu il} ,
\end{equation}
where summation over repeated indices is assumed, indices $\mu, \nu \in$ $\{1,2,3,4\}$ with indices $\{1,2,3\}$ ( = $\{x,y,z\}$) labeling a right-handed rectilinear spatial reference frame and 4 ( =~$t$) labeling the time component, and $i,j \in \{1,2,\ldots,n\}.$ 

A well known example of  vector matrices are the Dirac gamma matrices found in relativistic wave equations for spin 1/2 particles such as the electron. Equation (\ref{DvD}) is then a special case of Pauli's Fundamental Theorem.$^{ \cite{pauli}}$ 

Commutation rules follow directly from (\ref{DvD}). For example, let $\Lambda = $ $R$ be a rotation of the $x,y$-plane about the origin through an infinitesimal angle $\theta.$ Then $\cos{\theta} \approx 1$ and $\sin{\theta} \approx \theta$ and ${\Lambda_{\mu}}^{\nu}V_{\nu } \approx$ $ \{V_x - \theta V_y, V_y + \theta V_x,V_z,V_t\},$ where indices $ij$ can be inferred from the context and are dropped. Now let $J_{z}$ be the $z$-angular momentum matrix which generates $R$ so that $D^{\pm}_{ij}(R) \approx$ $ I_{ij} \pm i \theta J_{zij},$ where $I_{ij}$ is the $n \times n$ unit matrix. Substituting these in (\ref{DvD}) gives
\begin{equation} \label{DvD2} [J_{z}, V_{x}]_{ik} = J_{zij} V_{xjk}- V_{xij} J_{zjk} = i V_{yik} ,
\end{equation}
$$ [J_{z}, V_{y}]_{ik} = J_{zij} V_{yjk}- V_{yij} J_{zjk} = - i V_{xik} .$$
Matrix multiplication and summation over repeated indices are understood. These two commutation rules and the commutation rules like it can be used to calculate the vector matrices $V_{\mu ij}$ of any given $n \times n$ matrix representation of the homogeneous Lorentz group. 

Since there are three angular momentum components, $J_{i}$ with $i \in \{1,2,3\} ,$ and four vector indices, $V_{\mu}$ with $\mu \in \{1,2,3,4\},$ there are twelve commutation rules like (\ref{DvD2}). Another twelve come from the boost generators $K_{i},$ for a total of 24 commutation rules to be satisfied by vector matrices. Often, given a set of matrices $J_{i}$ and $K_{i},$ the commutation rules have only trivial, null vector matrix solutions.  

Irreducible Lorentz representations can be characterized by the spin pair $(A,B),$ with $2A$ and $2B$ integers. We show that nontrivial vector matrices do not exist for irreducible Lorentz representations. 

However, there are nonzero vector matrices for reducible representations with spin \linebreak $(A,B) \oplus (C,D)$ when $\mid A -C \mid $ = $ 1/2$ and $\mid B - D \mid $ = $ 1/2.$ The results for more general spin types can be deduced from the results obtained here. Explicit formulas for vector matrix components are collected in an Appendix which are to be used with the standard set of angular momentum and boost generator matrices  displayed in Sec.~\ref{Lorentz}

Momentum generates translations. Translations, rotations and boosts together make up the inhomogeneous Lorentz group of spacetime symmetries known as the Poincar\'{e} group. Translations commute, so the matrix representation of momentum $P_{\mu ij}$ is a set of four vector matrices  that commute. Once the expressions for vector matrices are found it is but a short step to apply the rule $[P_{\mu},P_{\nu}]$ = 0 and obtain an $n \times n$ matrix representation of the Poincar\'{e} algebra. 

The representations of the Poincar\'{e} algebra found here are nonunitary, being extensions of nonunitary $n \times n$ matrix representations of the homogeneous Lorentz group and should not be confused with the unitary representations of the Poincar\'{e} algebra familiar since the work by Wigner$.^{ \cite{unitary1} - \cite{unitary4}}$
 
Sections ~\ref{Poincare} and \ref{Lorentz} establish notation. In Sec.~\ref{Poincare} the well-known commutation rules of the Poincar\'{e} algebra are presented,$^{ \cite{genref0}- \cite{genref4}}$ and  the rules for vector matrices are identified. In Sec.~\ref{Lorentz} formulas for the matrices of the irreducible matrix representations $(A,B)$ of the (homogeneous) Lorentz group are recalled from the literature.$^{ \cite{standardL}}$ The details of the derivation of the vector matrices take up most of the remainder of the article. In Sec.~\ref{4-VectorsI} we find that there are no vector matrices with any of the irreducible matrix representations of the Lorentz group, so, in Sec.~\ref{4-VectorsII} we consider reducible Lorentz representations with spin $(A,B)\oplus(C,D).$  In Sections~\ref{4-VectorsI} and \ref{4-VectorsII} we take advantage of the diagonal nature of matrices $J_z$ and $K_z.$ The step-up and step-down matrices $J^{\pm}$ give recursion relations in Sec.~\ref{4-VectorsIII}. Then the recursion relations are solved in Sec.~\ref{4-VectorsIV} for matrices $V^{\pm}$ which are combinations of $V_x$ and $V_y.$ The results determine $V_z$ and $V_t$ in Sec.~\ref{4-VectorsV}. Having found expressions for vector matrices, momentum matrices are obtained in Sec.~\ref{momentum}.   An Appendix collects expressions for the vector matrix components in terms of the spins $A,B,C,D$ and two free parameters $t_{AB}^{12}$ and $t_{CD}^{21}.$

The derivation illustrates standard techniques by first applying commutation rules with diagonal matrices and then dealing with step-up and step-down matrices. Also potentially interesting to students is the formulation and display of the matrices for spin $(A,B)\oplus(C,D)$. The formulas for the components of vector matrices in Appendix A serve as reference for more advanced students and practitioners.  

The derivation here is based on the Poincar\'{e} algebra. Many others have deduced formulas for vector matrices in studies of first order relativistically invariant field equations $\beta^{\mu}_{ik} \partial_{\mu} \psi_{k}$ = $\psi_{k},$ where we use `$\beta$' for the vector matrices since that is a common notation for vector matrices in such investigations. Concise expressions for vector matrices have been obtained by Lyubarskii \cite{Lbskii} and are recalled in Appendix B for convenience.

A problem set is attached to help assure understanding, to provide the reward of investigation, and to offer an exploration of topics extending beyond the text.

\section{The Poincar\'{e} Commutation Rules } \label{Poincare}  

There are ten Poincar\'{e} generators, the four components of momentum generate translations, the three components of angular momentum generate rotations and the three components of boost generators generate boosts. The commutator of any two different Poincar\'{e} generators is a homogeneous linear combination of generators.  The commutation rules of the Poincar\'{e} algebra are well known$.^{ \cite{genref0} - \cite{genref4}}$

The fifteen rules that do not involve the momentum form the commutation rules of the homogeneous Lorentz algebra,  
\begin{equation} \label{jj} [J_i,J_j] = i \epsilon_{ijk} J_k \hspace{1cm} [J_i,K_j] = i \epsilon_{ijk} K_k  \hspace{1cm} [K_i,K_j] = - i \epsilon_{ijk} J_k \end{equation}
where $J_k$ is the $k$ component of the angular momentum, $K_k$ is the $k$ component of the boost generator and $\epsilon_{ijk}$ determines the sign: $\epsilon_{ijk}$ is antisymmetric in $ijk$  with $\epsilon_{xyz} = 1$ and $i,j,k \in \{x,y,z\}.$

Twenty-four rules are linear and homogeneous in momentum components. They can be deduced from (\ref{DvD}) by appropriate choices for $\Lambda.$ A `vector matrix' is a set of four square matrices that obeys the 24 rules linear in momentum. Writing the four components of a vector as $V_\mu$ = $\{ V_i,V_t\},$ $\mu \in \{x,y,z,t\},$ the rules are
\begin{equation} \label{jv} [J_i,V_j] = i \epsilon_{ijk} V_k  \hspace{1cm} [K_i,V_j] =  -i  V_t \delta_{ij} \end{equation}
\begin{equation} \label{jv4} [J_i,V_t] = 0  \hspace{1cm} [K_i,V_t] =  -i V_i \end{equation}
References \cite{unitary3} and \cite{unitary4} display rules using the same spacetime metric as here, the diagonal matrix with diagonal $\{1,1,1,-1\},$ but with contravariant vector matrices $\{ V^i,V^t\}$ = $\{ V_i,-V_t\}.$ This accounts for minus signs in the rules involving $V_t.$ 

The final six rules are quadratic in momentum components and state that momentum components $P_\mu$ commute. Thus the $P_{\mu}$ obey the rules (\ref{jv}) and (\ref{jv4}) for vectors as well as the following six rules that say that the components commute,
\begin{equation} \label{pp} [P_\mu,P_\nu] = 0 \end{equation}
where $\mu, \nu \in \{x,y,z,t\}.$

Eqns. (\ref{jj}) - (\ref{pp}) (with $P_\mu$ for $V_\mu $ in (\ref{jv}) and (\ref{jv4})) are the commutation rules of the Poincar\'{e} algebra. 

\section{Irreducible Representations of the Lorentz Group } \label{Lorentz}

Matrices representing the components of angular momentum and boosts of the homogeneous Lorentz group and satisfying (\ref{jj}) are well known and given by standard expressions.$^{ \cite{standardL}}$ The irreducible representations  of the (homogeneous) Lorentz group are cataloged by the pair $(A,B),$ where 2A and 2B are non-negative integers. Following standard procedures$^{ \cite{standardL},\cite{weinberg2}},$ one begins by writing a $2A+1$ dimensional representation of the rotation group:
\begin{equation} \label{J+} M^{+ (A)}_{\sigma \sigma_1} = (M^{(A)}_{x} + i M^{(A)}_{y})_{\sigma \sigma_1} = r^{(A)}_{\sigma_1} \delta_{\sigma , \sigma_1 + 1}  \end{equation}
\begin{equation} \label{J-} M^{- (A)}_{\sigma \sigma_1} = (M^{(A)}_{x} - i M^{(A)}_{y})_{\sigma \sigma_1} = s^{(A)}_{\sigma_1} \delta_{\sigma , \sigma_1 - 1} \end{equation}
\begin{equation} \label{Jz} (M^{(A)}_{z})_{\sigma \sigma_1} =  \sigma_1 \delta_{\sigma  \sigma_1}   \end{equation}
where 
\begin{equation} \label{sigma} \sigma, \sigma_1 \in \{-A,-A+1, \ldots , A-1, A\}   \end{equation}
and
\begin{equation} \label{r}  r^{(A)}_{\sigma_1} = \sqrt{(A - \sigma_1)(A + \sigma_1 + 1)} \end{equation}
\begin{equation} \label{s}  s^{(A)}_{\sigma_1} = r^{(A)}_{-\sigma_1} = \sqrt{(A + \sigma_1)(A - \sigma_1 + 1)} \end{equation}
with $r^{(A)}_{\sigma_1}$ and $s^{(A)}_{\sigma_1}$ zero outside the allowed range of $\sigma_1$ in (\ref{sigma}). The matrices $\{M^{(A)}_{x}, $ $M^{(A)}_{y},$ $M^{(A)}_{z}\}$ satisfy the commutation rules of the Lie algebra of the rotation group: $[M_i,M_j] =$ $ i \epsilon_{ijk} M_k.$

Next define the $(2A+1)(2B+1)$ dimensional matrices $A_k$ and $B_k$,
\begin{equation} \label{Ai} (A_{k})_{ab,a_1b_1} \equiv  (M_{k}^{(A)})_{a,a_1} \delta_{b, b_1} \qquad (B_{k})_{ab,a_1b_1} \equiv  (M_{k}^{(B)})_{b,b_1} \delta_{a,a_1} \end{equation}

Then the $(A,B)$ representation of the Lorentz group can be taken to be the following $(2A+1)(2B+1)$ dimensional matrices $J_{k}^{(A,B)}$ and $K_k^{(A,B)}$ 
\begin{equation} \label{Ji} J_{k}^{(A,B)} = A_{k}+B_{k} \qquad  K_{k}^{(A,B)} = -i(A_{k}-B_{k}) \end{equation}
By (\ref{J+})-(\ref{Ji}), we find
\begin{equation} \label{JAB+} (J^{+ (A,B)})_{ab,a_1b_1} =  r^{(A)}_{a_1} \delta_{a,a_1+1} \delta_{b, b_1} + r^{(B)}_{b_1} \delta_{a,a_1} \delta_{b, b_1 +1}    \end{equation}
\begin{equation} \label{JAB-} (J^{- (A,B)})_{ab,a_1b_1} =  s^{(A)}_{a_1} \delta_{a,a_1-1} \delta_{b, b_1} + s^{(B)}_{b_1} \delta_{a,a_1} \delta_{b, b_1 -1}    \end{equation}
\begin{equation} \label{KAB+} (K^{+ (A,B)})_{ab,a_1b_1} = -i( r^{(A)}_{a_1} \delta_{a,a_1+1} \delta_{b, b_1} - r^{(B)}_{b_1} \delta_{a,a_1} \delta_{b, b_1 +1} )   \end{equation}
\begin{equation} \label{KAB-} (K^{- (A,B)})_{ab,a_1b_1} = -i( s^{(A)}_{a_1} \delta_{a,a_1-1} \delta_{b, b_1} - s^{(B)}_{b_1} \delta_{a,a_1} \delta_{b, b_1 -1} )   \end{equation}
\begin{equation} \label{JABz} (J_{z}^{(A,B)})_{ab,a_1b_1} = (a + b) \delta_{a,a_1} \delta_{b, b_1}    \end{equation}
\begin{equation} \label{KABz} (K_{z}^{(A,B)})_{ab,a_1b_1} = -i(a - b) \delta_{a,a_1} \delta_{b, b_1}  \end{equation}
We often suppress spin labels or indices. Note that $J_z$ and $K_z$ are diagonal; their nonzero components have the double index $ab$ equal to the double index $a_1b_1.$

\pagebreak
\section{Irreducible Lorentz Reps - No Solutions } \label{4-VectorsI}

Suppose we choose as angular momentum and boost matrices the $J_{i}$ and $K_i$ of the $(A,B)$ irreducible representation of the Lorentz group. One can check that the commutation rules (\ref{jj}) are satisfied. 

Next, we seek vector matrices $V_{\mu}$ that solve the 24 commutation rules (\ref{jv}) and (\ref{jv4}). Let us begin with the rules (\ref{jv}) and (\ref{jv4}) that involve the diagonal matrices $J_z$ and $K_z.$ By (\ref{jv}), we have $[J_z,V_x] = i V_y$ and $[J_z,V_y] = -i V_x$ which imply that $V_x = [J_z,[J_z,V_x]]$. Hence
\begin{equation} \label{jzjzpx} V_x = [J_z^{(A,B)},[J_z^{(A,B)},V_x]] = (J_z^{(A,B)})^2 V_x -2J_z^{(A,B)} V_x J_z^{(A,B)} +V_x (J_z^{(A,B)})^2 \end{equation}
We also have $[ K_z,V_x] = 0,$ which implies
\begin{equation} \label{kzpx} [ K_z^{(A,B)},V_x] = K_z^{(A,B)} V_x - V_x K_z^{(A,B)} = 0 \end{equation}
With (\ref{JABz}) and (\ref{KABz}) for $J_z^{(A,B)}$ and $K_z^{(A,B)}$, we get from (\ref{jzjzpx})
\begin{equation} \label{jpx} (V_x)_{ab,a_1b_1} = [(a+b)-(a_1+b_1)]^2 (V_x)_{ab,a_1b_1} \end{equation}
and from (\ref{kzpx}) we get
\begin{equation} \label{kpx}  [(a-b)-(a_1-b_1)] (V_x)_{ab,a_1b_1} = 0 . \end{equation}
One can quickly show that (\ref{jpx}) and (\ref{kpx}) imply that either
\begin{equation} \label{kpx2} a - a_1 = b - b_1 = \pm \frac{1}{2} \hspace{1cm} {\mathrm{or}} \hspace{1cm}(V_x)_{ab,a_1b_1} = 0 \end{equation}
But, by (\ref{sigma}), the indices $a$ and $a_1$ must differ by an integer $n$ since $a$ and $a_1$ range over the set $\{-A,-A+1,\ldots,A-1,A\}.$ Since the indices cannot differ by $\pm1/2$, we get  the trivial result,
\begin{equation} \label{kpx3} a-a_1 = b-b_1  = n \neq \pm \frac{1}{2}\hspace{1cm} \Rightarrow \hspace{1cm} (V_x)_{ab,a_1b_1} = 0 .  \end{equation}
Thus $V_x$ vanishes and that implies that all four $V_\mu$ vanish because the others can be written as commutators of $V_x$ with $J_z^{+ (A,B)},$ $J_y^{+ (A,B)},$ and $K_x^{+ (A,B)}$ by (\ref{jv}). We have
\begin{equation} \label{kpx4} \hspace{3cm}V_{\mu} = 0 . \hspace{2cm} ({\mathrm{Irreducible}} \quad J^{(A,B)}, \enskip K^{(A,B)}) \end{equation}

There are no non-zero vector matrices $V_\mu,$ and therefore also no non-zero momentum matrices $P_\mu,$ satisfying the Poincar\'{e} commutation rules with the angular momentum and boost matrices of an irreducible representation of the homogeneous Lorentz group.

\section{Reducible Lorentz Reps} \label{4-VectorsII}

Since irreducible representations fail to produce interesting results, we next try reducible representations of the Lorentz group. The simplest of these have angular momentum and boost matrices in block diagonal form with two irreducible representations along the diagonal. Consider the $(2A+1)(2B+1)$ + $(2C+1)(2D+1)$ dimensional representation of the Lorentz group which combines an $(A,B)$ and a $(C,D)$ representation. 

In block matrix notation, we have
\begin{equation} \label{2blockJ} J = \pmatrix{J_{11} & 0 \cr 0 & J_{22}} = \pmatrix{J^{(A,B)} & 0 \cr 0 & J^{(C,D)}} \qquad K = \pmatrix{K_{11} & 0 \cr 0 & K_{22}} =  \pmatrix{K^{(A,B)} & 0 \cr 0 & K^{(C,D)}} \end{equation}
with $J^{(A,B)}$ and $K^{(A,B)}$ as in Section \ref{Lorentz} and similar expressions for $J^{(C,D)}$ and $K^{(C,D)}.$ Putting the unknown vector matrices $V_{\mu}$ in block matrix form, we get 
\begin{equation} \label{2blockg} V_{\mu} = \pmatrix{V_{\mu 11} & V_{\mu 12} \cr V_{\mu 21} & V_{\mu 22}}  \end{equation}

The commutator $[J,V]$ in block matrix form is
\begin{equation} \label{comJg} [J,V] = \pmatrix{J_{11}.V_{11}-V_{11}.J_{11} & J_{11}.V_{12}-V_{12}.J_{22} \cr J_{22}.V_{21}-V_{21}.J_{11} & J_{22}.V_{22}-V_{22}.J_{22}}  \end{equation}
The commutation rules (\ref{jv}) and (\ref{jv4}), by inspection of the diagonal blocks of (\ref{comJg}), do not mix $(A,B)$ and $(C,D)$ representations.  Thus the results of Section \ref{4-VectorsI} for irreducible Lorentz representations apply to $V_{11}$ and $V_{22}$ and we infer that $V_{11}$ = 0 and $V_{22}$ = 0. However, the 12 and 21 blocks of $[J,V]$ do mix the (A,B) and the (C,D) representations and so the results of Section \ref{4-VectorsI} do not apply to the off-diagonal blocks of $V.$ Thus we have, by Section \ref{4-VectorsI} and without loss of generality,
\begin{equation} \label{offdiag} V_{\mu} = \pmatrix{0 & V_{\mu 12} \cr V_{\mu 21} & 0} .  \end{equation}
The unknowns are now the components of the off-diagonal blocks $V_{\mu 12}$ and $V_{\mu 21}$ of the four vector matrices $V_{\mu}.$ 

As in Section \ref{4-VectorsI}, we consider the equations in (\ref{jv}) and (\ref{jv4}) that involve the diagonal matrices $J_z$ and $K_z.$ Here, as there, we have $V_x = [J_z,[J_z,V_x]]$ from $[J_z,V_x] = i V_y$ and $[J_z,V_y] = -i V_x.$  For the 12 block, by (\ref{2blockJ}), (\ref{comJg}), and (\ref{offdiag}), these commutation rules give
\begin{equation} \label{jzjzpxABCD} V_{x12} = [J_z,[J_z,V_x]]_{12} = (J^{(A,B)}_z)^2 V_{x12} -2J^{(A,B)}_z V_{x12} J^{(C,D)}_z +V_{x12} (J^{(C,D)}_z)^2 \end{equation}
And we have  $[K_z,V_x] = 0,$ which gives
\begin{equation} \label{kzpx2} [K_z,V_x]_{12} = K^{(A,B)}_z.V_{x12}-V_{x12}.K^{(C,D)}_z = 0 . \end{equation}
By (\ref{JABz}) and (\ref{KABz}) for $J^{(A,B)}_z,$ $J^{(C,D)}_z,$ $K^{(A,B)}_z,$ and $K^{(C,D)}_z$ in (\ref{jzjzpxABCD}) and (\ref{kzpx2}) we get
\begin{equation} \label{jpxABCD} (V_{x12})_{ab,cd} = [(a+b)-(c+d)]^2 (V_{x12})_{ab,cd} \end{equation}
\begin{equation} \label{kpxABCD}  [(a-b)-(c-d)] (V_{x12})_{ab,cd} = 0 \end{equation}
By (\ref{jpxABCD}) and (\ref{kpxABCD}), we find that either
\begin{equation} \label{kpx2ABCD} a - c = b - d = \pm \frac{1}{2} \hspace{1cm} {\mathrm{or}} \hspace{1cm}(V_{x12})_{ab,cd} = 0 \end{equation}
Compare (\ref{kpx2ABCD}) with Eq. (\ref{kpx2}). In (\ref{kpx2}) $a$ and $a_1$ could not differ by a half, so there are no nontrivial solutions to (\ref{kpx2}). But $a$ and $c$ can differ by a half, so there may be nontrivial solutions to (\ref{kpx2ABCD}).

Since spins $A,B,C,D$ must be integers or half-integers and by (\ref{sigma}) $A-a,B-b,C-c,D-d$ are all integers, we find from (\ref{kpx2ABCD}) that when $V_{x12} \neq 0,$ 
\begin{equation} \label{ABCD} A = C + n + 1/2 \qquad B = D + m + 1/2,  \end{equation}
where $n$ and $m$ are integers. (Proof: In (\ref{kpx2ABCD}), let $a = A - n_1$ and $c = C - n_2,$ so that $A - C$ = $n_2-n_1 \pm 1/2.$) Thus either A or C is an integer while the other is half an odd integer and likewise for B and D. 

Using deltas to enforce (\ref{kpx2ABCD}), we have can write the $ab,cd$ component of the block $V_{x12}$ as
\begin{equation} \label{kpx3ABCD} (V_{x12})_{ab,cd} = t^{12}_{ab} \delta_{a,c+1/2} \delta_{b,d+1/2} + u^{12}_{ab}\delta_{a,c-1/2} \delta_{b,d-1/2}  , \end{equation} where $t^{12}_{ab}$ and $u^{12}_{ab}$ are as-yet-undetermined coefficients.

From $J_z$ and $V_x$ we can get $V_y;$ by (\ref{jv}), we have $iV_{y} =$ $[J_z,V_x].$ By (\ref{JABz}), (\ref{2blockJ}), (\ref{2blockg}), (\ref{comJg}) and (\ref{kpx3ABCD}), we get the 12 block of $iV_y,$
\begin{equation} \label{py3} i(V_{y12})_{ab,cd} = t^{12}_{ab} \delta_{a,c+1/2} \delta_{b,d+1/2} - u^{12}_{ab}\delta_{a,c-1/2} \delta_{b,d-1/2} \end{equation}

Comparing (\ref{kpx3ABCD}) and (\ref{py3}) suggest defining simpler matrices $V^+$ and $V^-$ by
\begin{equation} \label{p+1} (V^{+}_{12})_{ab,cd} \equiv \frac{1}{2}(V_{x12} + iV_{y12})_{ab,cd} = t^{12}_{ab} \delta_{a,c+1/2} \delta_{b,d+1/2}  \end{equation}
\begin{equation} \label{p-1} (V^{-}_{12})_{ab,cd} \equiv \frac{1}{2}(V_{x12} - iV_{y12})_{ab,cd} =  u^{12}_{ab}\delta_{a,c-1/2} \delta_{b,d-1/2} \end{equation}

To determine the 21-block of $V^{+}$ and $V^{-},$ exchange the $(A,B)$ and $(C,D)$ blocks in the matrices $J_k$ and $K_k,$ (\ref{2blockJ}). The effect is to exchange $a \leftrightarrow c,$ $b \leftrightarrow d,$ $1  \leftrightarrow 2$ in (\ref{p+1}) and (\ref{p-1}). We get
\begin{equation} \label{p+2} (V^{+}_{21})_{cd,ab} \equiv \frac{1}{2}(V_{x21} + iV_{y21})_{cd,ab} = t^{21}_{cd} \delta_{c,a+1/2} \delta_{d,b+1/2}  \end{equation}
\begin{equation} \label{p-2} (V^{-}_{21})_{cd,ab} \equiv \frac{1}{2}(V_{x21} - iV_{y21})_{cd,ab} =  u^{21}_{cd}\delta_{c,a-1/2} \delta_{d,b-1/2} \end{equation}
Thus we have
\begin{equation} \label{offdiag+} V^+ = \pmatrix{0 & V_{12}^+ \cr V_{21}^+ & 0} = \pmatrix{0 & t^{12}_{ab} \delta_{a,c+1/2} \delta_{b,d+1/2} \cr t^{21}_{cd} \delta_{c,a+1/2} \delta_{d,b+1/2} & 0} \end{equation}
and
\begin{equation} \label{offdiag-} V^- = \pmatrix{0 & V_{12}^- \cr V_{21}^- & 0} = \pmatrix{0 & u^{12}_{ab}\delta_{a,c-1/2} \delta_{b,d-1/2} \cr u^{21}_{cd}\delta_{c,a-1/2} \delta_{d,b-1/2} & 0}  \end{equation}

Note carefully that there may not be a variable $t^{12}_{ab}$ for every allowed value of $a$ and $b$ because the factors $\delta_{a,c+1/2}$ and $\delta_{b,d+1/2}$ in (\ref{p+1}). These factors imply that $(V^{+}_{12})_{ab,cd}$ depends on $t^{12}_{ab}$ only when $c$ = $a - 1/2$ and $d$ = $b - 1/2.$ Thus the indices of $t^{12}_{ab}$ are restricted also by the ranges of $c$ and $d$, forcing the indices $a$ and $b$ of $t^{12}_{ab}$ to be in the following ranges
\begin{equation} \label{abRanges} t^{(12)}_{ab}: \quad \max[-A,-C+ \frac{1}{2}] \leq a \leq \min[A,C+\frac{1}{2}]   \end{equation}
\begin{equation} \label{abRanges2} t^{(12)}_{ab}:  \quad \max[-B,-D+\frac{1}{2}] \leq b \leq \min[B,D+\frac{1}{2}]  \end{equation}
Similar restrictions apply to the ranges of the variables $t^{21}_{cd},$ $u^{12}_{ab},$ and $u^{21}_{cd}.$

In this section we have considered the commutation rules involving the vector matrices $V_{x}$ and $V_{y}$ and the diagonal matrices $J_z$ and $K_z.$ The resulting simplification is evident in Eqns. (\ref{offdiag+}) and (\ref{offdiag-}). Next we consider the commutation rules of $V_x$ and $V_y$ with the given matrices $J_x$ and $J_y.$

\section{Step-Up and Down; Recursions} \label{4-VectorsIII}

Define the `step-up' and `step-down' matrices $J^+$ and $J^-,$ 
\begin{equation} \label{Jpm} J^{+} = J_x + iJ_y  \hspace{1cm} {\mathrm{and}} \hspace{1cm} J^{-} = J_x - iJ_y,  \end{equation}
so-named by their effect on eigenvectors of $J_z.$ The commutation rules (\ref{jv}) applied to the commutators of $J^+$ and $J^-$ with the matrices $V^+$ and $V^-$ yield  
\begin{equation} \label{J+g+} [J^\pm,V^\pm] = 0  \hspace{1cm} {\mathrm{and}} \hspace{1cm} [J^\pm,V^\mp] = \pm V_z ,  \end{equation}
which are simpler than the equivalent equations with $J_x$ and $J_y.$

By (\ref{JAB+}), (\ref{JAB-}), and (\ref{2blockJ}), we find the needed formulas for $J^{\pm},$
\begin{equation} \label{2blockJ+} J^+  =  \pmatrix{r^{(A)}_{a_1} \delta_{a,a_1+1} \delta_{b, b_1} + r^{(B)}_{b_1} \delta_{a,a_1} \delta_{b, b_1 +1} & 0 \cr 0 & r^{(C)}_{c_1} \delta_{c,c_1+1} \delta_{d, d_1} + r^{(D)}_{d_1} \delta_{c,c_1} \delta_{d, d_1 +1}} \hspace{.5cm} \end{equation}
\begin{equation} \label{2blockJ-} J^-  =  \pmatrix{s^{(A)}_{a_1} \delta_{a,a_1-1} \delta_{b, b_1} + s^{(B)}_{b_1} \delta_{a,a_1} \delta_{b, b_1 -1} & 0 \cr 0 & s^{(C)}_{c_1} \delta_{c,c_1-1} \delta_{d, d_1} + s^{(D)}_{d_1} \delta_{c,c_1} \delta_{d, d_1 -1}} \end{equation}

Consider the 12-block $[J^+,V^+]_{12}$ = 0. With $V^+$ from  (\ref{offdiag+}) and $J^+$ from (\ref{2blockJ+}),  we have
\begin{equation} \label{ra} (r^{(A)}_{a-1} t^{12}_{a-1,b} - r^{(C)}_{c} t^{12}_{ab}) \delta_{a,c+3/2} \delta_{b,d+1/2} = 0  \end{equation}
\begin{equation} \label{rb}   (r^{(B)}_{b-1} t^{12}_{a,b-1} - r^{(D)}_{d} t^{12}_{ab}) \delta_{a,c+1/2} \delta_{b,d+3/2} = 0 \end{equation}
From $[J^-,V^-]_{12}$ = 0, with $V^-$ from  (\ref{offdiag-}) and $J^-$ from (\ref{2blockJ-}),  we get
\begin{equation} \label{sa} (s^{(A)}_{a+1} u^{12}_{a+1,b} - s^{(C)}_{c} u^{12}_{ab}) \delta_{c,a+3/2} \delta_{d,b+1/2} = 0 \hspace{1cm}\end{equation}
\begin{equation} \label{sb}  (s^{(B)}_{b+1} u^{12}_{a,b+1} - s^{(D)}_{d} u^{12}_{ab}) \delta_{c,a+1/2} \delta_{d,b+3/2} = 0  \end{equation}
Note that these are recursion relations in the indices $a$ and $b.$

From the right-hand equation in (\ref{J+g+}), we have $[J^+,V^-]_{12}$ = $-[J^-,V^+]_{12}.$ With $V^{\pm}$ and $J^{\pm}$ from (\ref{offdiag+}), (\ref{offdiag-}), (\ref{2blockJ+}) and (\ref{2blockJ-}), we have
\begin{equation} \label{rs1} (r^{(A)}_{a-1} u^{12}_{a-1,b} - r^{(C)}_{c} u^{12}_{ab} + s^{(B)}_{b+1} t^{12}_{a,b+1} - s^{(D)}_{d} t^{12}_{ab}) \delta_{a,c+1/2} \delta_{b,d-1/2} = 0  \hspace{1cm} \end{equation}
\begin{equation} \label{rs2} (r^{(B)}_{b-1} u^{12}_{a,b-1} - r^{(D)}_{d} u^{12}_{ab} + s^{(A)}_{a+1} t^{12}_{a+1,b} - s^{(C)}_{c} t^{12}_{ab}) \delta_{c,a+1/2} \delta_{d,b-1/2} = 0 \end{equation}

Without loss of generality we can assume that $A > C,$ because we can interchange the $AB$ and $CD$ blocks in the $J$ and $K$ matrices whenever $C > A.$ Thus, by interchanging the $AB$ and $CD$ blocks if necessary, we can assume that $n$ is positive or zero in (\ref{ABCD}). We have
\begin{equation} \label{AC}  
A = C + n + \frac{1}{2}  \hspace{1cm} (n \geq 0) . 
\end{equation}
The factor $\delta_{a,c+3/2}$ in Eq. (\ref{ra}) implies that the recursion relation is trivial unless $c$ = $a - 3/2$ = $-1/2 + a - 1$ = $C + n - A + a -1,$ where we use $-1/2$ = $C+n-A$ from (\ref{AC}). By definition (\ref{r}), $r^{(C)}_{c}$ vanishes for $c - C \geq 0,$ i.e. for $c - C = n - A + a -1 \geq 0.$ Thus $r^{(C)}_{c}$ vanishes for $a \geq$ $A-n+1$ and the recursion relation (\ref{ra}) implies that
\begin{equation} \label{ra1} (r^{(A)}_{a-1} t^{12}_{a-1,b} - r^{(C)}_{c} t^{12}_{ab}) \delta_{b,d+1/2} = r^{(A)}_{a-1} t^{12}_{a-1,b}  \delta_{b,d+1/2} = 0\hspace{1cm} (a \geq A-n+1) .  \end{equation}
By (\ref{abRanges}) and (\ref{ra1}) and since  $r^{(A)}_{a-1}$ is not zero for $A \geq a \geq -A+1,$ we have $t^{12}_{a-1,b}$ = 0 for $\min[A, C+1/2] \geq a \geq \max[A - n + 1,-A+1,-C+1/2].$ If $n \geq 1$, then $\min[A, C+1/2]$ = $A$ and $\max[A - n + 1,-A+1,-C+1/2] \leq A,$ implying $a = A$ is in the range and therefore $t^{12}_{A-1,b}$ = 0. The recursion formula (\ref{ra}) makes all the $t^{12}$s vanish. We have
\begin{equation} \label{ra4}  t^{12}_{-A+2,b} = \cdots = t^{12}_{A-1,b} = 0  \hspace{1cm} (A \geq C+3/2) \end{equation}

Substituting $t^{12}_{ab}$ = 0 into Eq. (\ref{rs1}), now gives an equation similar to (\ref{ra}), but with $u$s instead of $t$s. The same reasoning that gave (\ref{ra4}) gives all $u^{12}_{ab}$ = 0. And similar reasoning applied to the 21-block gives $t^{21}_{cd}$ = $u^{21}_{cd}$ = 0, and $V^+ = V^- = 0,$
 i.e. $V_x$ = $V_y$ = 0.
By (\ref{jv}), all four vector matrices vanish and we have, for $n \geq 1,$
\begin{equation} \label{g12}  V_{\mu} = 0  \hspace{1cm} (A \geq C+3/2) . \end{equation}
Since $n$ is positive or zero by assumption in (\ref{AC}), we conclude that  $V_{\mu} = 0$ unless $n$ = 0.

When $n$ = 0 in (\ref{AC}), there are no recursion relations that reduce to a single term as in (\ref{ra1}). Note that the requirement $a \geq A-n+1$ would require $a \geq A+1$ and that exceeds the range of index $a,$ by (\ref{sigma}). Thus there may be nonzero vector matrices for $A$ = $C + 1/2.$

Dropping the restriction that  $A$  be larger  than $C$ we see that $C$ = $A +1/2$ could give nonzero matrices $V_{\mu}.$ By investigations of the recursion relations for the $B$ and $D$ indices we find  that the vector matrices $V_{\mu}$ are zero unless
\begin{equation} \label{+-1/2}  A = C \pm \frac{1}{2} \hspace{1cm} {\mathrm{and}} \hspace{1cm} B = D \pm \frac{1}{2}\end{equation}
Thus the requirement that there be nonzero vector matrices severely limits the choice of spins $(A,B)\oplus(C,D).$

\section{Applying the Recursion Relations} \label{4-VectorsIV}

The problem of finding vector matrices $V_\mu$ has reduced to finding the quantities $t^{12}_{ab},$ $t^{21}_{cd},$ $u^{12}_{ab},$ and $u^{21}_{cd}$ in  $V^+$ and $V^-,$ (\ref{offdiag+}) and (\ref{offdiag-}). In this section we find that the recursion relations (\ref{ra})-(\ref{rs2}) allow the $t$s and $u$s to be written as functions of $A,B,C,D$ with two of the $t$s, $t^{12}_{AB}$ and $t^{21}_{CD},$ remaining as free parameters.

We need only consider the two possibilities in  (\ref{+-1/2}) with $A>C$, since we can interchange the $(A,B)$ and $(C,D)$ blocks of the $J$ and $K$ matrices if $C>A.$ The two cases are $A = C + 1/2; B = D \pm 1/2.$ We take $B>D$ as Case 1 and $D>B$ as Case 2.

Case 1: $A = C + 1/2; B = D + 1/2.$ Consider the recursion relation (\ref{ra}) and start with $a$ = $A.$ Then the delta function factor $\delta_{a,c+3/2}$ in (\ref{ra}) determines $c$ = $a-3/2$ = $A-3/2$ = $C-1.$ For any allowed index $b,$ by (\ref{r}) and (\ref{ra}), we have 
\begin{equation} \label{ra6}  t^{12}_{A-1,b} = \frac{r^{(C)}_{C-1}}{r^{(A)}_{A-1}} t^{12}_{A,b} = \frac{\sqrt{[C-(C-1)][C+(C-1)+1]}}{\sqrt{[A-(A-1)][A+(A-1)+1]}} t^{12}_{A,b} = \frac{\sqrt{2A-1}}{\sqrt{2A}} t^{12}_{A,b} .\end{equation}
Substituting successive values of $a,$ $a$ = $A-1,\ldots,-A+2,$ into the recursion relation, we find that
\begin{equation} \label{ra7}  t^{12}_{a,b} =  \frac{\sqrt{A+a}}{\sqrt{2A}} t^{12}_{A,b} \end{equation}
The recursion (\ref{rb}) gives similar results for the $b$ indices. We get
\begin{equation} \label{rab1}  t^{12}_{a,b} =  \frac{\sqrt{A+a}}{\sqrt{2A}}\frac{\sqrt{B+b}}{\sqrt{2B}} t^{12}_{A,B} . \hspace{2cm}({\mathrm{Case}} \enskip {\mathrm{1}})\end{equation}
for all values of the indices, $-A+1 \leq a \leq A$ and $ -B +1\leq b \leq B$ allowed by Eq. (\ref{abRanges}). 

Recursions (\ref{sa}) and (\ref{sb}) give a formula for the $u$s. We get
\begin{equation} \label{sab1}  u^{12}_{a,b} =  \frac{\sqrt{A-a}}{\sqrt{2A}} \frac{\sqrt{B-b}}{\sqrt{2B}}u^{12}_{-A,-B} . \hspace{2cm}({\mathrm{Case}} \enskip {\mathrm{1}})\end{equation}
for all allowed values of the indices, $-A \leq a \leq A-1$ and $ -B \leq b \leq B-1.$ 

Substituting $t^{12}_{ab}$ and $u^{12}_{ab}$ from (\ref{rab1}) and (\ref{sab1}) into Eq.(\ref{rs1}), we get
\begin{equation} \label{rsab1}  \frac{\sqrt{A+a}}{\sqrt{2A}}\frac{\sqrt{B-b}}{\sqrt{2B}}( t^{12}_{A,B}  + u^{12}_{-A,-B}) = 0 .\end{equation}
For $a \neq -A$ and $b \neq B$ in (\ref{rsab1}), we have
\begin{equation} \label{rsab2}   \hspace{3cm}  u^{12}_{-A,-B}  = - t^{12}_{A,B}  \hspace{2cm}({\mathrm{Case}} \enskip {\mathrm{1}})\end{equation}
Eq. (\ref{rs2}) produces the same result. Equations (\ref{rab1}), (\ref{sab1}) and (\ref{rsab2}) determine the quantities $t^{12}_{ab}$ and $u^{12}_{ab}$ for Case 1, $A=C+1/2$ and $B=D+1/2.$ 

Case 2: $A$ = $C + 1/2;$ $B$ = $D - 1/2.$ Start the recursion relation (\ref{rb}) with $b$ = $B.$ The delta function factor $\delta_{b,d+3/2}$ in (\ref{rb}) determines $d$ = $b - 3/2$ = $B - 3/2$ = $D-2.$ Then for any allowed index $a,$ by (\ref{r}) and (\ref{rb}) we have
\begin{equation} \label{rb2}  t^{12}_{a,B-1} = \frac{r^{(D)}_{D-2}}{r^{(B)}_{B-1}} t^{12}_{a,B} = \frac{\sqrt{[D-(D-2)][D+(D-2)+1]}}{\sqrt{[B-(B-1)][B+(B-1)+1]}} t^{12}_{a,B} = \sqrt{2}  \enskip  t^{12}_{a,B} \end{equation}
Substituting successive values of $b,$ $b$ = $B-1,\ldots,-B+1,$ into the recursion relation, we find that
\begin{equation} \label{rb3}  t^{12}_{a,b} =  \sqrt{B-b+1} \enskip t^{12}_{a,B} . \end{equation}
Applying the recursion (\ref{ra7}) for the $a$-index, we get
\begin{equation} \label{rb4}  t^{12}_{a,b} =  \frac{\sqrt{A+a}}{\sqrt{2A}}  \sqrt{B-b+1} \enskip  t^{12}_{A,B} .\hspace{2cm}({\mathrm{Case}} \enskip {\mathrm{2}}) \end{equation}
for all allowed indices, $-A+1 \leq a \leq A$ and $-B \leq b \leq B$ in Eq. (\ref{abRanges}).

Similar work with Eq. (\ref{sa}) and Eq. (\ref{sb}) gives a formula for the $u$s. We get
\begin{equation} \label{sab12}  u^{12}_{a,b} =  \frac{\sqrt{A-a}}{\sqrt{2A}} \sqrt{B+b+1} \enskip  u^{12}_{-A,-B} \hspace{2cm}({\mathrm{Case}} \enskip {\mathrm{2}})\end{equation}
for all allowed indices, $-A \leq a \leq A-1$ and $-B \leq b \leq B.$  

Substituting $t^{12}_{ab}$ and $u^{12}_{ab}$ from (\ref{rb4}) and (\ref{sab12}) in Eqns.(\ref{rs1}) and (\ref{rs2}), we get
\begin{equation} \label{rsab3}    \hspace{3cm}   u^{12}_{-A,-B}  =  t^{12}_{A,B}  \hspace{2cm}({\mathrm{Case}} \enskip {\mathrm{2}}).\end{equation}
Equations (\ref{rb4}), (\ref{sab12}) and (\ref{rsab3}) determine the quantities $t^{12}_{ab}$ and $u^{12}_{ab}$ for Case 2, $A=C+1/2$ and $B=D-1/2.$ 

To determine the 21-block quantities $t^{21}_{cd}$ and $u^{21}_{cd},$ interchange the $(A,B)$ and $(C,D)$ blocks in the $J_k$ and $K_k$ matrices (\ref{2blockJ}). The effect is to exchange $a \leftrightarrow c,$ $b \leftrightarrow d,$ $1  \leftrightarrow 2,$ Case 1 $\leftrightarrow$ Case 4, and Case 2 $\leftrightarrow$ Case 3. Thus the equations for $t^{21}_{cd}$ and $u^{21}_{cd}$ can be transcribed from  (\ref{rab1}), (\ref{sab1}) and (\ref{rsab2}) for Case 1 and from (\ref{rb4}), (\ref{sab12}) and (\ref{rsab3}) for Case 2.

At this point we could write expressions for the vector matrices $V^+$ and $V^-.$ However, we wait until we can also write $V_z$ and $V_t.$

\section{Completing the Set of Vector Matrices} \label{4-VectorsV}

Expressions for vector matrices $ V_z$ and $ V_t$ in terms of $V^+$ and $V^-$ follow from the commutation rules. By (\ref{J+g+}), $ V_z$ = $[J^+,V^-]$ and, with $J^+$ and $V^-$ from (\ref{offdiag-}) and (\ref{2blockJ+}), we have
\begin{equation} \label{gz} V_{z12} = \frac{1}{2}[(r^A_{a-1} u^{12}_{a-1,b} - r^C_c u^{12}_{a,b}) \delta_{a,c+1/2} \delta_{b,d-1/2} +(r^B_{b-1} u^{12}_{a,b-1} - r^D_d u^{12}_{a,b}) \delta_{a,c-1/2} \delta_{b,d+1/2}].\end{equation}
The formula expresses the components of the matrix $V_z$ in terms of known functions.

An expression for $ V_t$ can be found by (\ref{jv}), $[iK^+,V^-]$ = $i[K_x + iK_y, V_x - i V_y]/2$ = $ V_t.$ By (\ref{KAB+}) and (\ref{2blockJ}), the needed matrix $i K^+$ is found to be
\begin{equation} \label{2blockK+} i K^+  =  \pmatrix{r^{(A)}_{a_1} \delta_{a,a_1+1} \delta_{b, b_1} - r^{(B)}_{b_1} \delta_{a,a_1} \delta_{b, b_1 +1} & 0 \cr 0 & r^{(C)}_{c_1} \delta_{c,c_1+1} \delta_{d, d_1} - r^{(D)}_{d_1} \delta_{c,c_1} \delta_{d, d_1 +1}} \hspace{.5cm}  \end{equation}
This together with $V^{-}$ from (\ref{offdiag-}) gives
\begin{equation} \label{gt} V_{t12} = \frac{1}{2}[(r^A_{a-1} u^{12}_{a-1,b} - r^C_c u^{12}_{a,b}) \delta_{a,c+1/2} \delta_{b,d-1/2} -(r^B_{b-1} u^{12}_{a,b-1} - r^D_d u^{12}_{a,b}) \delta_{a,c-1/2} \delta_{b,d+1/2}].\end{equation}
To obtain the 21-blocks of $V_z$ and $V_t$ make the following exchanges in the 12-blocks: $ 1 \leftrightarrow 2,$ $ \{A,a\} \leftrightarrow \{C,c\},$ and $ \{B,b\} \leftrightarrow \{D,d\}$

Note that the terms in brackets in these expressions for $ V_z$ and $ V_t$ differ only in sign, so it is preferable to display the combinations $ V_z \pm$ $ V_t$ since the many cancelations simplify the displayed expressions.  

The explicit formulas for the matrix components of $V_{\mu}$ are lengthy so they have been placed in an appendix. The collection of formulas in Appendix A is the main result of this paper.

\section{Momentum Matrices} \label{momentum}

Momentum matrices $P_{\mu}$ satisfy the commutation rules (\ref{jv}) and (\ref{jv4}) for vector matrices as well as rule (\ref{pp}) which says that momentum matrices commute, $[P_{\mu},P_{\nu}]$ = 0. 

For Case 1, with $A = C + 1/2; B = D + 1/2$ and by (\ref{pp+12}) - (\ref{ppz21}), the 11-block of $[P^+,P^-]$ gives 
\begin{equation} \label{PP1}  [P^+,P^-]_{11;a,a_1,b,b_1} =  -\frac{Ab+aB}{\sqrt{AB}} \enskip  t^{12}_{AB} t^{21}_{CD} \delta_{a,a_1} \delta_{b,b_1} =0, \hspace{0.5cm}(A = C + 1/2; \enskip B = D + 1/2) \end{equation}
where $P^{\pm}$ = $(P_{x} \pm iP_{y})/2.$ Choosing $a$ = $A$ and $b$ = $B,$ we get    
\begin{equation} \label{PP2}  [P^+,P^-]_{11;A,a_1,B,b_1} =  -2\sqrt{AB} \enskip  t^{12}_{AB} t^{21}_{CD} \delta_{A,a_1} \delta_{B,b_1} = 0.  \end{equation}
Since neither $A$ nor $B$ can be zero in Case 1 and the delta functions are nonzero for $a_1$ = $A$ and $b_1$ = $B,$ we have
\begin{equation} \label{PP3}   t^{12}_{AB} = 0   \hspace{1cm} {\mathrm{or}} \hspace{1cm}   t^{21}_{CD} = 0.  \end{equation}

Inspection of (\ref{pp+12}) - (\ref{mmz21}) shows that $t^{12}_{AB}$ is a common factor in all components of $P_{12}^{\mu}$ and $t^{21}_{CD}$ is a common factor in all components of $P_{21}^{\mu}.$ Thus, by (\ref{PP3}), the momentum matrix has one of the following two block matrix forms, either
\begin{equation} \label{PP4}   P_{\mu} = \pmatrix{0 && 0 \cr V_{\mu}^{21} && 0}  \hspace{1cm} {\mathrm{or}} \hspace{1cm} P_{\mu} = \pmatrix{0 && V_{\mu}^{12} \cr 0 && 0}   .  \end{equation}
One can show that the same result holds for all cases, $A = C \pm 1/2 ;$ $ B = D \pm 1/2$: the momentum matrices $P_{\mu}$ are the vector matrices $V_{\mu},$ (\ref{pp+12})-(\ref{mmz21}), with either the 12-block or the 21-block set equal to zero.

\appendix

\section{Formulas for Vector Matrices } \label{Formulas}

We now give formulas for the matrix components of $V_{\mu}$ for spin $(A,B) \oplus (C,D).$ Results for $V_{\mu}$ in the general case of spin $ \sum (A_i,B_i)$ can be deduced from the formulas here, see Problem 4. 

The angular momentum matrices $J_i$ and the boost matrices $K_i$ can be found by (\ref{2blockJ}) and (\ref{JAB+}) through (\ref{KABz}).

The matrices $V_{\mu}$ are $n \times n$ square matrices with $n$ = $(2A+1)(2B+1)+(2C+1)(2D+1),$ with components put in block matrix form,
\begin{equation} \label{appBLOCK} V = \pmatrix{ (V_{11})_{ab,a1b1} && (V_{12})_{ab,cd} \cr (V_{21})_{cd,ab} && (V_{22})_{cd,c1d1}}  = \pmatrix{ 0 && (V_{12})_{ab,cd} \cr (V_{21})_{cd,ab} && 0} ,\end{equation}
since the 11 and 22 blocks are null in all four cases.  The indices $\{ab,cd\}$ and $\{cd,ab\}$are pairs of double indices originating from the $(A,B) \oplus (C,D)$ representation of the Lorentz group.   Replacing the pairs of double indices with pairs of single indices, e.g. $\{ab,cd\} \rightarrow \{i,j\},$ is discussed in the problem set, see Problem 3.
The indices have ranges $a \in \{-A, -A+1, \ldots, A \},$  $b \in \{-B, -B+1, \ldots, B \},$ $c \in \{-C, -C+1, \ldots, C \},$ $d \in \{-D, -D+1, \ldots, D \}.$

The formulas for the $(V_{12}^{\pm})_{ab,cd}$ blocks for Case 1, $A = C + 1/2; B = D + 1/2,$ come from substituting (\ref{rab1}), (\ref{sab1}) and (\ref{rsab2}) in (\ref{p+1}) and (\ref{p+2}). The $(V_{z12} \pm V_{t12})_{ab,cd}$ blocks for Case 1 come from substituting (\ref{sab1}) and (\ref{rsab2}) in (\ref{gz}) and (\ref{gt}). For Case 2, $A = C + 1/2; B = D - 1/2,$ use (\ref{rb4}), (\ref{sab12}) and (\ref{rsab3}) instead of (\ref{rab1}), (\ref{sab1}) and (\ref{rsab2}). 

Cases 1 and 2  become Cases 4 and 3 when we swap $(A,B)\oplus(C,D) \rightarrow $  $(C,D)\oplus(A,B).$ Thus the formulas of Case 4 and 3 can be obtained from  those of Cases 1 and 2 by the exchanges $\{A,a\} \leftrightarrow \{C,c\},$ $\{B,b\} \leftrightarrow \{D,d\},$ $12  \leftrightarrow 21.$

Four cases give nontrivial solutions, $A = C \pm 1/2;$ $B = D \pm 1/2.$

Case 1: $A = C + 1/2; B = D + 1/2.$

\begin{equation} \label{pp+12} (V^{\pm}_{12})_{ab,cd} =  \frac{1}{2}(V_{x12} \pm iV_{y12})_{ab,cd} = \pm \frac{\sqrt{A \pm a}}{\sqrt{2A}}\frac{\sqrt{B \pm b}}{\sqrt{2B}} \enskip t^{12}_{AB} \delta_{a,c \pm 1/2} \delta_{b,d \pm 1/2}  \end{equation}
\begin{equation} \label{ppz12}  \frac{1}{2}(V_{z12} \pm V_{t12})_{ab,cd} =   - \frac{\sqrt{A \pm a}}{\sqrt{2A}}\frac{\sqrt{B \mp b}}{\sqrt{2B}} \enskip t^{12}_{AB}\delta_{a,c \pm 1/2} \delta_{b,d \mp 1/2} \end{equation}
\begin{equation} \label{pp+21} (V^{\pm}_{21})_{cd,ab} =  \frac{1}{2}(V_{x21} \pm iV_{y21})_{cd,ab} =  \pm \sqrt{A \mp a}\sqrt{B \mp b} \enskip t^{21}_{CD} \delta_{c,a \pm 1/2} \delta_{d,b \pm 1/2} , \end{equation}
\begin{equation} \label{ppz21}  \frac{1}{2}(V_{z21} \pm V_{t21})_{cd,ab} = + \sqrt{A \mp a}\sqrt{B \pm b} \enskip t^{21}_{CD}\delta_{c,a \pm 1/2} \delta_{d,b \mp 1/2} \quad     \end{equation}


Case 2: $A = C + 1/2; B = D - 1/2.$

\begin{equation} \label{pm+12} (V^{\pm}_{12})_{ab,cd} =  \frac{1}{2}(V_{x12} \pm iV_{y12})_{ab,cd} =  \frac{\sqrt{A \pm a}}{\sqrt{2A}}\sqrt{D \mp d} \enskip t^{12}_{AB} \delta_{a,c \pm 1/2} \delta_{b,d \pm 1/2}  \end{equation}
\begin{equation} \label{pmz12}   \frac{1}{2}(V_{z12} \pm V_{t12})_{ab,cd} =  \pm \frac{\sqrt{A \pm a}}{\sqrt{2A}}\sqrt{D \pm d} \enskip t^{12}_{AB}\delta_{a,c \pm 1/2} \delta_{b,d \mp 1/2}    \end{equation}
\begin{equation} \label{pm+21} (V^{\pm}_{21})_{cd,ab} =  \frac{1}{2}(V_{x21} \pm iV_{y21})_{cd,ab} =   \sqrt{A \mp a}\frac{\sqrt{D \pm d}}{\sqrt{2D}} \enskip t^{21}_{CD} \delta_{c,a \pm 1/2} \delta_{d,b \pm 1/2}  \end{equation}
\begin{equation} \label{pmz21}  \frac{1}{2}(V_{z21} \pm V_{t21})_{cd,ab} = \mp \sqrt{A \mp a} \frac{\sqrt{D \mp d}}{\sqrt{2D}} \enskip t^{21}_{CD}\delta_{c,a \pm 1/2} \delta_{d,b \mp 1/2} \end{equation}

Case 3: $A = C - 1/2; B = D + 1/2.$
\begin{equation} \label{mp+12} (V^{\pm}_{12})_{ab,cd} =  \frac{1}{2}(V_{x12} \pm iV_{y12})_{ab,cd} =    \sqrt{C \mp c}\frac{\sqrt{B \pm b}}{\sqrt{2B}} \enskip t^{12}_{AB} \delta_{a,c \pm 1/2} \delta_{b,d \pm 1/2}  \end{equation}
\begin{equation} \label{mpz12}   \frac{1}{2}(V_{z12} \pm V_{t12})_{ab,cd} =  \mp \sqrt{C \mp c}\frac{\sqrt{B \mp b}}{\sqrt{2B}} \enskip t^{12}_{AB}\delta_{a,c \pm 1/2} \delta_{b,d \mp 1/2} \end{equation}
\begin{equation} \label{mp+21} (V^{\pm}_{21})_{cd,ab} =  \frac{1}{2}(V_{x21} \pm iV_{y21})_{cd,ab} =    \frac{\sqrt{C \pm c}}{\sqrt{2C}}\sqrt{B \mp b} \enskip t^{21}_{CD} \delta_{c,a \pm 1/2} \delta_{d,b \pm 1/2}  \end{equation}
\begin{equation} \label{mpz21}  \frac{1}{2}(V_{z21} \pm V_{t21})_{cd,ab} =  \pm \frac{\sqrt{C \pm c}}{\sqrt{2C}}\sqrt{B \pm b} \enskip t^{21}_{CD}\delta_{c,a \pm 1/2} \delta_{d,b \mp 1/2}  \end{equation}

Case 4: $A = C - 1/2; B = D - 1/2.$
\begin{equation} \label{mm+12} (V^{\pm}_{12})_{ab,cd} =  \frac{1}{2}(V_{x12} \pm iV_{y12})_{ab,cd} =   \pm \sqrt{C \mp c}\sqrt{D \mp d} \enskip t^{12}_{AB} \delta_{a,c \pm 1/2} \delta_{b,d \pm 1/2}  \end{equation}
\begin{equation} \label{mmz12}  \frac{1}{2}(V_{z12} \pm V_{t12})_{ab,cd} =   + \sqrt{C \mp c}\sqrt{D \pm d} \enskip t^{12}_{AB}\delta_{a,c \pm 1/2} \delta_{b,d \mp 1/2}   \end{equation}
\begin{equation} \label{mm+21} (V^{\pm}_{21})_{cd,ab} =  \frac{1}{2}(V_{x21} \pm iV_{y21})_{cd,ab} =  \pm \frac{\sqrt{C \pm c}}{\sqrt{2C}}\frac{\sqrt{D \pm d}}{\sqrt{2D}} \enskip t^{21}_{CD} \delta_{c,a \pm 1/2} \delta_{d,b \pm 1/2}  \end{equation}
\begin{equation} \label{mmz21}  \frac{1}{2}(V_{z21} \pm V_{t21})_{cd,ab} =  - \frac{\sqrt{C \pm c}}{\sqrt{2C}}\frac{\sqrt{D \mp d}}{\sqrt{2D}} \enskip t^{21}_{CD}\delta_{c,a \pm 1/2} \delta_{d,b \mp 1/2}  \end{equation}

\section{The Lyubarskii Formulas } \label{CGC}

The study of invariant wave equations of first order began with the Dirac equation. Since these wave equations involve vector matrices and since there is and has been considerable interest in such wave equations, there are many approaches containing many varied expressions for vector matrices. The literature is too vast to cite here. 

One thing that seems to be missing from these activities is the recognition that a set of commuting vector matrices act as the momentum generators of translations that complete, with the angular momentum matrices generators, a set of generators satisfying the Poincar\'{e} algebra. 

However that may be, Lyubarskii has obtained remarkably compact expressions for the formulas that take up so much space in Appendix \ref{Formulas}. 
Lyubarskii's approach to vector matrices is paraphrased here.  See Ref. \cite{Lbskii} for details. 

In the wave equation $$ \beta^{\mu}_{ik} \partial_{\mu} \psi_{k} = \psi_{i} $$ the partial $\partial_{\mu}$ transforms as a four-vector, i.e. with spin $(1/2,1/2).$ The four vector index $\mu$ can thereby be replaced with a double index $ m n,$ with $m$ and $n$ having values $+1/2$ or $-1/2.$ Likewise the index $i$ for the components $\psi_{k} $ that transform with spin $(A,B)$ can be replaced by a double index $ab,$ $\psi_{k}$ = $\psi_{ab}$. The wave equation requires $\psi_{i}$ = $\psi_{cd}$ to be a linear combination of $$ P^{mn}_{cd;ab} \partial_{mn} \psi_{ab} \, , $$  where $cd$ is the double index for those components of $\psi_{i}$ with spin $(C,D)$ and $P^{mn}_{cd;ab}$ is simply related to  the vector matrices $\beta^{\mu}_{ik}.$ The notation $P^{mn}_{cd;ab}$ mimics Lyubarskii's notation.

Sums of basic vectors for $\partial_{mn} \psi_{ab},$ i.e. the direct product $(1/2,1/2)\times(A,B),$  transform like the basic vectors of $\psi_{cd}$ with spin $(C,D)$ when the coefficients $ P^{mn}_{cd;ab}$ are Clebsch-Gordan coefficients, $$ P^{mn}_{cd;ab} = \langle 1/2 \,  m, A a \mid C c \rangle \langle 1/2 \,  n, B b \mid D d \rangle \, .  $$ Clebsch-Gordan coefficients are also called vector addition coefficients. The basic properties of Clebsch-Gordan coefficients imply that the coefficients are zero unless the spins are related by $C$ = $A \pm 1/2$ and $D$ = $B \pm 1/2.$

One needs to sum over the spin $(1/2,1/2)$ indices to get a four-vector, $$ (\beta_{\mu 21})_{cd,ab} = \lambda_{21}({\bar{V}}_{\mu })_{m0,0n} \, P^{mn}_{cd;ab} = \lambda_{21} ({\bar{V}}_{\mu })_{m0,0n} \langle 1/2 \, m, A a \mid C c \rangle \langle 1/2 \,  n, B b \mid D d \rangle \, , $$   where ${\bar{V}}_{\mu}$ is the vector matrix for spin $(1/2,0)\oplus(0,1/2),$ $\lambda_{21}$ is an arbitrary constant and $m$ and $n$ are summed over $\pm 1/2.$ The choice of the matrix ${\bar{V}}_{\mu}$ must be coordinated with the representation of the angular momentum and boost matrices. 

For the standard representations in Sec. \ref{Lorentz}, one may choose $$ {\bar{V}}_{x} = \pmatrix{0 && 1 \cr 1 && 0} \quad {\bar{V}}_{y} = \pmatrix{0 && i \cr -i && 0} \quad {\bar{V}}_{z} = \pmatrix{-1 && 0 \cr 0 && 1} \quad {\bar{V}}_{t} = \pmatrix{-1 && 0 \cr 0 && -1} \, ,$$ with indices $$ mn = \pmatrix{-\frac{1}{2} -\frac{1}{2}&& -\frac{1}{2} +\frac{1}{2} \cr +\frac{1}{2} -\frac{1}{2} && +\frac{1}{2} +\frac{1}{2}} \, . $$

For the $\beta_{(12)}$ blocks, one has $$ (\beta_{\mu 12})_{ab,cd} =  \lambda_{12}({\bar{V}}_{\mu })_{m0,0n} \langle 1/2 \,  m, C c \mid A a \rangle \langle 1/2 \,  n, D d \mid B b \rangle \, . $$ 
Substituting $ \beta_{\mu 12}$ and $ \beta_{\mu 21}$ for the blocks $V_{\mu 12}$ and $V_{\mu 21}$ in (\ref{offdiag}) yield vector matrices. These together with the $J$s and $K$s of (\ref{2blockJ}) satisfy the commutation rules of the Poincar\'{e} algebra displayed in Section \ref{Poincare}.

One can show that the expressions derived in this paper and displayed in Appendix \ref{Formulas} are equivalent to the formulas obtained by Lyubarskii and displayed in this Appendix. The collective knowledge of Clebsch-Gordan coefficients makes the Lyubarskii formulas attractive.

\section{Problems} 

\noindent 1. Use Appendix A with $(A,B)$ = $(1/2,0)$ and $(C,D)$ = $(0,1/2)$ and an appropriate choice of free parameters $t^{12}_{AB}$ and $t^{21}_{CD}$ to obtain a set of Dirac gamma matrices $V_{\mu}$ = $\gamma_{\mu}.$ Show that the matrices satisfy $V_{\mu} \cdot V_{\nu}+V_{\nu} \cdot V_{\mu}$ = $2\eta_{\mu \nu} I,$ where $\eta$ is the metric and $I$ is the 4x4 unit matrix. (The matrix indices on the $V_{\mu}$ and $I$ are suppressed.)

\vspace{0.3cm}
\noindent 2. The Euclidean group in two dimensions, $E_2$.$^{\cite{tungE2}}$ Rotations $R$ in the $x,y$ plane about the origin can be represented with the $z$-angular momentum generator $J_z$ given by $$ J_z = \pmatrix{0 && -i && 0 \cr i && 0 &&0 \cr 0 && 0 && 0},   $$ which acts on vector components $\{x,y,1\}.$ The constant third component indicates the direct product of $\{x,y\}$ with a scalar which doesn't change under rotations. Let $V_{xij}$ and $V_{yij}$ be vector matrices obeying the commutation rules $[J_z,V_x]_{ij} =$ $iV_{yij}$ and $[J_z,V_y]_{ij} =$ $-iV_{xij}.$ The group of motions in the $x,y$-plane including the rotations $R$ and $x$ and $y$ translations is called $E_2.$  (a) Find  $V_{xij}$ and $V_{yij}$ from the commutation rules. (b) Find the momentum matrices $P_{xij}$ and $P_{yij}.$ 

\vspace{0.3cm}
\noindent 3. Single index notation. One may reduce the double index $(a,b)$ to a single index $i.$ For example, with $A$ = $B$ = 1/2, we can write $(a,b) \rightarrow i$ by $\{ (-1/2,-1/2)\rightarrow 1,(-1/2,+1/2)\rightarrow 2,(+1/2,-1/2)\rightarrow 3,(+1/2,+1/2)\rightarrow 4 \} ,$ so that $i \in $ $\{1,2,3,4\}.$ With this numbering system and the formulas of Case 3 in Appendix A, find the ten $7 \times 7$ matrices $J_k,$ $K_k,$ $V_{\mu}$ for spin $(A,B) \oplus (C,D)$ = $(1/2,1/2) \oplus (1,0).$ Verify one commutation rule from each set of rules (\ref{jj}) - (\ref{pp}).

\vspace{0.3cm}
\noindent 4. What spins $A,B,C,D,E,F$ give non-zero vector matrices $V_{\mu}$ when the $J$s and $K$s form the $(A,B) \oplus (C,D) \oplus (E,F)$ reducible  representation of the Lorentz algebra (\ref{jj})? Argue from the linearity of $V_{\mu}$ in rules (\ref{jv}) and (\ref{jv4}) or, as in Sec.~\ref{4-VectorsI}, use the diagonal matrices $J_z$ and $K_z$ to decide which blocks of the vector matrices are possibly non-zero or argue otherwise.

\vspace{0.3cm}
\noindent 5. `Vector Representation.' Calculate the momentum matrices $P_{\mu}$ for $(A,B)$ = $(1/2,1/2)$ and $(C,D)$ = $(0,0)$ with $P_{\mu}^{12} \neq 0$ and $P_{\mu}^{21} = 0.$ Find the similarity transformation $S$ that takes these $P_{\mu}$s to the following set of matrices, $SPS^{-1} =$ $P^{ \prime},$ where
$$ P_{\mu}^{ \prime} = \pmatrix{0 && 0 && 0 && 0 && \delta_{\mu,x} \cr 0 && 0 && 0 && 0 && \delta_{\mu,y} \cr 0 && 0 && 0 && 0 && \delta_{\mu,z} \cr 0 && 0 && 0 && 0 && \delta_{\mu,t} \cr 0 && 0 && 0 && 0 && 0} .$$
Show that $$ (1+i x_{\mu} P_{\mu}^{ \prime})^2 = (1 + i 2 x_{\mu} P_{\mu}^{ \prime}) $$ and $$ e^{i x_{\mu} P_{\mu}^{ \prime}} = 1+i x_{\mu} P_{\mu}^{ \prime} .$$ 

\vspace{0.3cm}
\noindent 6. Now show that the new generators in the previous problem (the $J^{ \prime}$s, $K^{ \prime}$s and $P^{ \prime}$s) produce matrices $D(L,a)$ that represent Poincar\'{e} transformations in the form 
$$ D(L,a) = \pmatrix{L^{\mu}_{\nu} && a^{\mu} \cr 0  && 1}, $$
where $L_{\mu,\nu}$ is a (4x4) homogeneous Lorentz transformation and $a$ = $\{a_x,a_y,a_z,a_t\}$ is the translation vector.$^{\cite{hamermesh2}}$ (First, verify that the AB block of $J^{ \prime}$s and $K^{ \prime}$s form the regular representation for which the matrices $D(L,0)$ are well known. Then use the off-diagonal $P^{ \prime}$s from Problem 6 to finish. Or proceed some other way.)

\vspace{0.3cm}
\noindent 7. Show the vector matrices for each case in Appendix \ref{Formulas} can be made equal to those in Appendix \ref{CGC}. The matrices are equal when the constant $\lambda_{12}$ is the constant $t^{12}_{AB}$ times a function of spins $A$ and $B$ and $\lambda_{21}$ is $t^{21}_{CD}$ times a function of spins $C$ and $D.$ [For Case 2, \linebreak $A$ = $C$ $+1/2$ and $B$ = $D - 1/2,$ one finds that $V_{\mu 12}$ = $\beta_{\mu 12}$ when $\lambda_{12}$ = $  \sqrt{2B+1}\, t^{12}_{AB}\, .$]

\vspace{0.3cm}
\noindent 8. Find all spins $(A,B) \oplus (C,D)$ such that the commutators of the vector matrices are proportional to the angular momentum and boost generators expressed in four dimensional notation, i.e.
$$ [V^{\mu},V^{\nu}] = k J^{\mu \nu}, $$
where $k$ is constant, $J^{\mu \nu}$ is antisymmetric in $\mu, \nu,$ $J^{ij}$ = $\epsilon_{ijk} J_k,$ and $J^{it}$ = $K_i.$ [By number crunching I find there are just two solutions, $(0,1/2) \oplus (1/2,0)$ and $(1/2,0) \oplus (0,1/2),$ i.e. the Dirac gamma matrices of Problem 1. This does not prove that there are no other solutions.]

\vspace{0.3cm}
\noindent 9. Find all spins $(A,B) \oplus (C,D)$ such that the anti-commutator of the vector matrices is proportional to the metric, that is
$$ \{V^{\mu},V^{\nu}\} = V^{\mu} \cdot V^{\nu}+ V^{\nu} \cdot V^{\mu} = k \eta^{\mu \nu}I, $$
where $k$ is constant, $\eta^{\mu \nu}$ = diag$\{-1,-1,-1,+1\}$ is the metric and $I$ is the unit matrix. [Again, the Dirac gamma matrices of Problem 1 are the only vector matrices that I have found to work. Can you show that the Dirac gammas are the only vector matrices  for spin $(A,B)\oplus (C,D)$ whose commutators and anticommutators produce the angular momentum matrices and the metric as in problems 8 and 9?]



\begin{thebibliography}{15}

\bibitem{pauli} J.J.Sakurai, Advanced Quantum Mechanics(Addison-Wesley, Reading, 1967), Appendix C (Pauli's fundamental theorem).



\bibitem{unitary1}  Wigner, E., Annals of Mathematics, 40, 149 (1939). 

\bibitem{unitary2} Kim, Y. S. and Noz, M. E., Theory and Applications of the Poincar\'{e} Group (D. Reidel Publishing Co., Dordrecht, Holland, 1986), Chapter III.

\bibitem{unitary3} Tung, W., Group Theory in Physics (World Scientific Publishing Co. Pte. Ltd., Singapore, 1985), Sec.~10.4. 

\bibitem{unitary4} Weinberg, S., The Quantum Theory of Fields, Vol. I (Cambridge University Press, Cambridge, 1995), Sec.~2.5 and references therein.



\bibitem{genref0} The Resource Letter in Ref. \cite{genref3} contains many references to articles published prior to 1981. References \cite{genref1} to \cite{genref4} are only some of the many resources available to those interested in the Poincar\'{e} group and its algebra.

\bibitem{genref1} Kim, Y. S. and Noz, M. E., op. cit.  

\bibitem{genref2} Tung, W., op. cit., Chapt.~10;

\bibitem{genref3} The Resource Letter: Stuewer, R. H., Am. J. Phys., Vol.~49, No.~4, April 1981, pp.~304-319.

\bibitem{genref4} Hamermesh, M., Group Theory and its Application to Physical Problems (Addison-Wesley Publishing Co., Reading, Massachusetts, 1964).

\bibitem{standardL} Weinberg, S., op. cit., Sec.~5.6 and references therein.

\bibitem{Lbskii} Lyubarskii, G. Ya., transl. by S. Dedijer, The Applications of Group Theory in Physics (Pergamon Press, Oxford, 1960), Chap. XVI.

\bibitem{weinberg2} Tung, W., op. cit., Sec.~10.3.

\bibitem{tungE2} Tung, W., op. cit., Sec.~9.1;

\bibitem{hamermesh2} Hamermesh, M., op. cit., Sec.~12.4.



 
\end{thebibliography}
\end{document}